\documentclass[conference, letterpaper, twocolumn]{IEEEtran}
\IEEEoverridecommandlockouts
\usepackage[table]{xcolor}
\usepackage[T1]{fontenc}
\usepackage{cite}
\usepackage{amsmath,amssymb,amsfonts}
\usepackage{algorithmic}
\usepackage{graphicx}
\usepackage{textcomp}
\usepackage{xcolor}
\usepackage{listings}
\usepackage{array}
\usepackage{float}
\usepackage{subcaption}
\usepackage{color}
\usepackage{booktabs}
\usepackage{tikz}
\usepackage{inconsolata}

\usetikzlibrary{tikzmark,calc}
\usetikzlibrary{arrows.meta}

\def\BibTeX{{\rm B\kern-.05em{\sc i\kern-.025em b}\kern-.08em
    T\kern-.1667em\lower.7ex\hbox{E}\kern-.125emX}}

\usepackage[absolute]{textpos}
\begin{textblock}{5}(11.8,0.8)
\end{textblock}

\begin{document}

\IEEEoverridecommandlockouts
\IEEEpubid{\makebox[\columnwidth]{ 979-8-3503-6378-4/24\$31.00 \copyright2024 IEEE \hfill} \hspace{\columnsep}\makebox[\columnwidth]{ }}

\title{SPRING: Systematic Profiling of Randomly Interconnected Neural Networks Generated by HLS\\
\thanks{This work was partially supported by DoE ASRSP GCFA Grant ID: SP0062070 and NSF POSE Phase II Award 2303700.}
}

\author{\IEEEauthorblockN{
Rui Shi, 
Seda Ogrenci}
\IEEEauthorblockA{Department of Electrical and Computer Engineering, Northwestern University
}
}

\maketitle
\begingroup\renewcommand\thefootnote{\textsection}
\endgroup

\begin{abstract}

Profiling is important for performance optimization by providing real-time observations and measurements of important parameters of hardware execution. Existing profiling tools for High-Level Synthesis (HLS) IPs running on FPGAs are far less mature compared with those developed for fixed CPU and GPU architectures and they still lag behind mainly due to their dynamic architecture. This limitation is reflected in the typical approach of extracting monitoring signals off of an FPGA device individually from dedicated ports, using one BRAM per signal for temporary information storage, or embedding vendor specific primitives to manually analyze the waveform. In this paper, we propose a systematic profiling method tailored to the dynamic nature of FPGA systems, particularly suitable for streaming accelerators. Instead of relying on signal extraction, the proposed profiling stream flows alongside the actual data, dynamically splitting and merging in synchrony with the data stream, and is ultimately directed to the processing system (PS) side. We conducted a preliminary evaluation of this method on randomly interconnected neural networks (RINNs) using the FIFO fullness metric, with co-simulation results for validation. 



\end{abstract}


\section{Introduction}



Machine learning applications are gainly immense popularity for their ability to predict complex tasks based on large amounts of collected data. Because they have specific structures that usually involve large amounts of matrix multiplications, they are mostly suited for platforms with custom architectures like GPUs. However, FPGAs also stand out for their flexibility in customizing architectures for applications deployed at the edge. One important architecture is the streaming architecture, which eliminates complex control logic and reduces unnecessary data movement back and forth. These architectures have promising applications in domains such as experimental scientific instrumentation \cite{10035147, 10596538, 10.3389/fdata.2022.787421} and high-performance imaging\cite{HLS4ML}\cite{blott2018finn}. However, unlike CPUs and GPUs, which feature dedicated profiling tools with built-in hardware performance counters for real-time performance and resource monitoring, FPGA applications generated by HLS flows lack a corresponding systematic profiling mechanism. This makes real-time analysis and diagnosis of machine learning workloads synthesized with HLS on FPGA-based systems challenging.


One way to profile the internal state of HLS streaming accelerators is to expose signals of interest to the top-level function and observe them using a vendor specific IP such as AMD's Integrated Logic Analyzer (ILA) core, or to use on-chip buffers to temporarily store monitored data, typically one BRAM per signal followed by periodic transfer of the collected information to a central unit or host for later analysis. While this method can effectively capture relevant signal activity, it involves significant manual effort, and the number of signals that can be monitored is constrained by available communication ports, on-chip memory resources, and the complexity of managing concurrency in streaming architectures\cite{goeders2016signal}\cite{choi2017hlscope}\cite{caba2020fpga}. These limitations significantly restrict the scope of profiling. Introducing specific pragmas to lower portions of the design from C++ to HDL for more fine-grained control  can help mitigate some of these challenges, but such solutions often require modifications at the compiler level\cite{RealProbe}. Few existing works focus on systematically collecting on-board runtime data from FPGAs or providing datasets that capture this information, despite its importance for understanding post-implementation behavior of HLS-generated designs and enabling future, more targeted optimizations.


\begin{figure}[h]
   \centering
   \includegraphics[width=\linewidth]{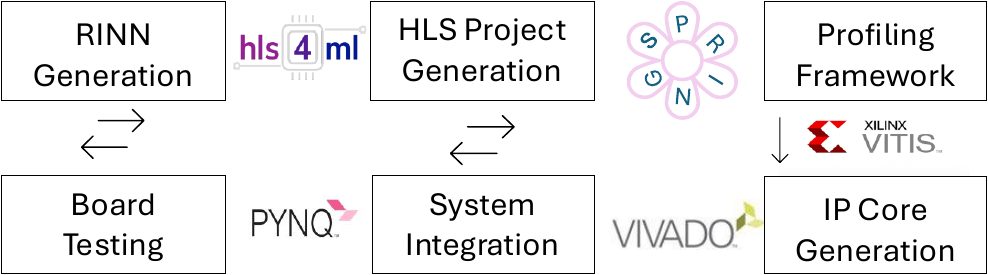}
   \caption{Profiling Flow}
   \label{fig:flow} 
\end{figure}

\begin{figure*}[ht]
    \centering
    \includegraphics[width=\linewidth]{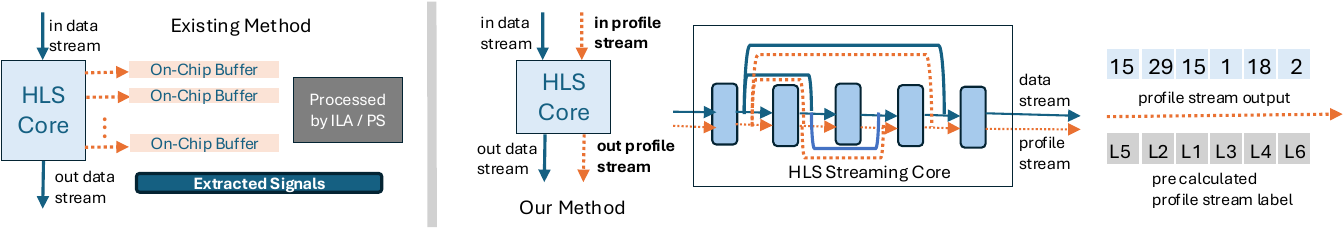}
    \caption{Comparison of the typical profiling structure and our proposed profiling structure.} 
    \label{fig:profiling_overview}
\end{figure*}

In this paper, we introduce a new systematic approach for profiling streaming machine learning accelerators purely at the HLS level, applied to a set of randomly interconnected neural networks (RINNs). We have constructed a fully automated system which generates RINNs and integrates them into an HLS design flow including the open source tool \texttt{hls4ml} \cite{Aarrestad_2021} in combination with a commercial FPGA implementation flow to generate a large volume of status data collected from completely synthesized and FPGA mapped benchmarks. Our main contributions are as follows:

\begin{itemize}
    \item A novel profiling framework capable of concurrently monitoring over 200 internal signals per design purely at the HLS level, with a focus on FIFO size metrics.
    
    \item Seamless integration of the on-FPGA profiling framework into the open-source tool \texttt{hls4ml}, enabling large-scale generation of RINN-based benchmark datasets.
    
    \item Analysis of neural network patterns and their impact on FIFO sizing, offering design guidance for initial parameter tuning in accelerator development.
\end{itemize}





The remainder of the paper is organized as follows. Section~\ref{method} describes our proposed method, the RINN architectures used, and the automated process for experimental setup. Section~\ref{result} presents the resource overhead, compares profiling results with cosimulation FIFO size, and explores FIFO size patterns across different RINN configurations. Finally, Section~\ref{conclusion} summarizes the paper and discusses current limitations and directions for future work.

\section{Dynamic On-FPGA Profiling Methodology} \label{method}

In this section, we first introduce a streaming profiling method and then apply this method to observe real-time FIFO fullness. 
Accurately monitoring FIFO fullness during runtime is important not only for designing resource-efficient streaming accelerators but also for ensuring their stable operation and avoiding deadlocks. Different FPGA systems might further introduce unpredictability to the utilization of FIFOs due to external factors such as the interaction of datapath blocks with external memory and timing imbalances. These impacts are challenging for HLS tools to prdict accurately. 

The dynamic profiling methodology we outline in the following has been integrated into \texttt{hls4ml} and it is also supported by a framework which generates randomly interconnected neural network benchmarks that can be automatically synthesized by Vitis HLS. A corresponding one-click system is developed to automate the entire process, from massive RINN benchmark generation to on-board data collection.

\subsection{Systematic Profiling Structures}

As shown in Fig. \ref{fig:profiling_overview}, the existing profiling method for HLS cores typically extracts signals one by one, each buffered by on-chip storage and then fed into an ILA for real-time waveform viewing or post-processing on the PS side of the FPGA device. In contrast, our proposed method generates a profiling data stream that flows alongside the actual data stream. Each time the data stream splits or merges, the profiling stream is split or merged accordingly.

\begin{minipage}{0.9\linewidth}

\begin{lstlisting}[caption={Streaming Profiling function}, label=lst:profile_code, basicstyle=\footnotesize\ttfamily, captionpos=b]
template <class pf_in_T, class pf_out_T, ...> 
void f(hls::stream<pf_in_T> &pf_in, 
hls::stream<pf_out_T> &pf_out, ...) {
  unsigned max_depth = 0;
  ...
  io_section: {
    #pragma HLS protocol fixed
    unsigned ffsize = data.size();
    if (max_depth < ffsize) max_depth = ffsize;
    in_data = data.read();
  }
  ...
  pf_out_T pf_out_data;
    pf_in_T pf_in_data = pf_in.read();
    for (int i = 0; i < pf_in_T::size; i++) {
        #pragma HLS UNROLL
        pf_out_data[i] = pf_in_data[i];
    }
    pf_out_data[pf_out_T::size-1] = max_depth;
    pf_out.write(pf_out_data);
}
\end{lstlisting}

\end{minipage}

\begin{minipage}{\linewidth}
\begin{lstlisting}[caption={Synthesized HDL Code}, label=lst:HDL_code, basicstyle=\footnotesize\ttfamily, captionpos=b]
...
// stream_fifo.v: mOutPtr
// mOutPtr is extracted out for data.size() 
always @(posedge clk) begin
    if (reset)
        mOutPtr <= {ADDR_WIDTH+1{1'b1}};
    else if (push & ~pop)
        mOutPtr <= mOutPtr + 1'b1;
    else if (~push & pop)
        mOutPtr <= mOutPtr - 1'b1;
end
...
//profiled_module.v (add function)
assign pf_out_din = {{{{{{pf_in2_ffsize}, 
{pf_in1_ffsize}}, {pf_in1_dout}}, {pf_in2_dout}};
...
//The profiling mechanism introduces an extra state
//for output data and profile data write,
// and they influence each other
parameter    ap_ST_fsm_state3 = 3'd4;
...
always @ (*) begin
 ap_block_state3 = ((pf_out_full_n==1'b0) |
 (pf_in1_empty_n==1'b0)|(pf_in2_cpy2_empty_n==1'b0)|
 (data_out_full_n==1'b0));
end


\end{lstlisting}
\end{minipage}

When the profiling stream passes through a module, the module first reads the incoming profiling data from the previous stage and appends the newly collected data to the end of the stream. In the case of a stream split, for example, when the current module has one input stream and multiple output streams as in a clone function, all current profiling data is written to the first output stream, while the second output stream is initialized with a placeholder value. For a stream merge, the first input stream is read and written to the output stream first, followed by the second input stream. The example profiling code is provided in Listing.\ref{lst:profile_code}, utilizing the size function \cite{forelli2024high}. The synthesized HDL code is provided in Listing.\ref{lst:HDL_code} for reference.

Further optimizations are possible, such as balancing the lengths of split profiling streams to reduce resource usage, or creating shortcuts to directly forward sufficiently long profiling streams to the dataflow’s final merging module while inserting a new placeholder at their original location. Once these stream-handling policies are defined, a predetermined output profiling label list can be generated. When the profiling stream is later retrieved from the IP’s output, the semantic meaning of each segment in the stream can be easily interpreted.

\subsection{Randomly Interconnected Neural Network} 
We implemented a simplified Keras version of the model based on a previously proposed system \cite{xie2019exploring}. We refer to the neural network described in this fashion as a Randomly Interconnected Neural Network (RINN) and we created an automated generator for RINNs based on the framework provided by \cite{geiss2023emulating}. A similar automation has been attempted in the past \cite{kuramochi2021low}, however, it was not applied to benchmark generation for FPGA synthesis and it was unrelated to profiling.


Since our focus is on the hardware aspects, particularly the FIFO sizing, we symbolically train the RINNs using the MNIST dataset. The original input format consists of 16 elements, and the output has 5 elements. To adapt this for our use case, we first pass the input through a dense layer to generate an arbitrary number of inputs as needed. We then apply a reshape operation to convert it into a shape of $(x,x,1)$. After that, we stack multiple Conv2D layers with the same shape to introduce random connectivity. Finally, the data flows through a flatten layer, followed by a dense layer with a sigmoid activation to produce a 5 elements output compatible with the MNIST dataset.

We also experimented with RINNs composed mainly of Dense layers, Concatenation, Add, ReLU, and Sigmoid. Based on observations from both co-simulation and profiling, the FIFO size is consistently no larger than 1. Examples of the generated RINNs can be found in Fig.~\ref{fig:complexity}. We are going to support more layer types in the near future for various RINNs.

\subsection{Automatic Profiling Flow on SoC FPGAs with an Example Metric}
To automate the entire process as shown in Fig. \ref{fig:flow}, we made the following efforts. Starting with a RINN generation script with extendable tunable connection density, layer size, and layer type \cite{geiss2023emulating}, we first extend \texttt{hls4ml} to support arbitrary amounts of inputs or outputs for the clone and merge functions. Presently, \texttt{hls4ml} does not support Vitis HLS accelerators on FPGA SoCs like PYNQ-Z2 or ZCU102, so we extended it and incorporated it into the PYNQ framework.
We create a profiling framework that can be applied after the \texttt{hls4ml} project is generated. The framework automatically parses the neural network information and selects the type of layers which need to be profiled with selective profile precision, where merging and splitting layers must be included. The parsing process saves the predetermined output profile label for result analysis. After the update, the new \texttt{hls4ml} project has two input streams and two output streams on its IP core interface. One is for the original data flow and the other is for the profiling data stream. Since the structure is fixed outside the HLS IP core, we create templates from Vivado integration to PYNQ execution scripts for the updated HLS IP core, and all of them are automatically applied by the profiling framework. Finally, an overall program coordinates the entire flow, including RINN generation, \texttt{hls4ml} project creation, profiling injection, Vitis HLS IP generation, Vivado system integration, PYNQ deployment, and post-execution data processing, enabling full automation with a single command.

\section{Results} \label{result}

This section consists of three parts. The first part analyzes the resource overhead introduced by the proposed profiling framework. The second part compares the co-simulation results with the profiling outputs to validate accuracy. The third part presents example profiling results across various RINN generation strategies.



\subsection{Resource Overhead Analysis of the Profiling Framework}

\begin{figure}[ht]
    \centering
    \includegraphics[width=\linewidth]{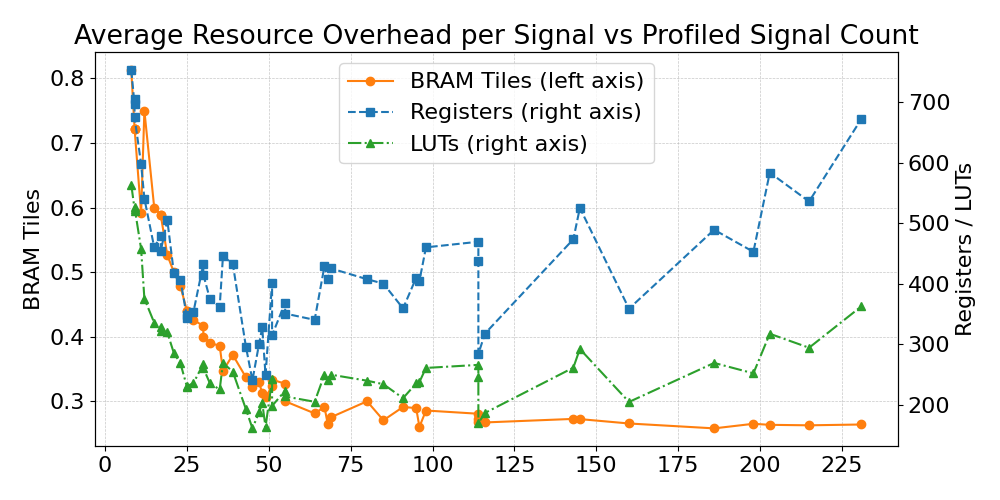}
    \caption{ZCU102, Conv2D stacking.  Data is collected from the Vivado implementation report, with the overhead computed by subtracting the resource requirement of the original (non-profiled) version from the design with profiling structures, and then averaged by the number of concurrently profiled signals.}
    \label{fig:resourceoverhead1}
\end{figure}

\begin{figure}[ht]
    \centering
    \includegraphics[width=\linewidth]{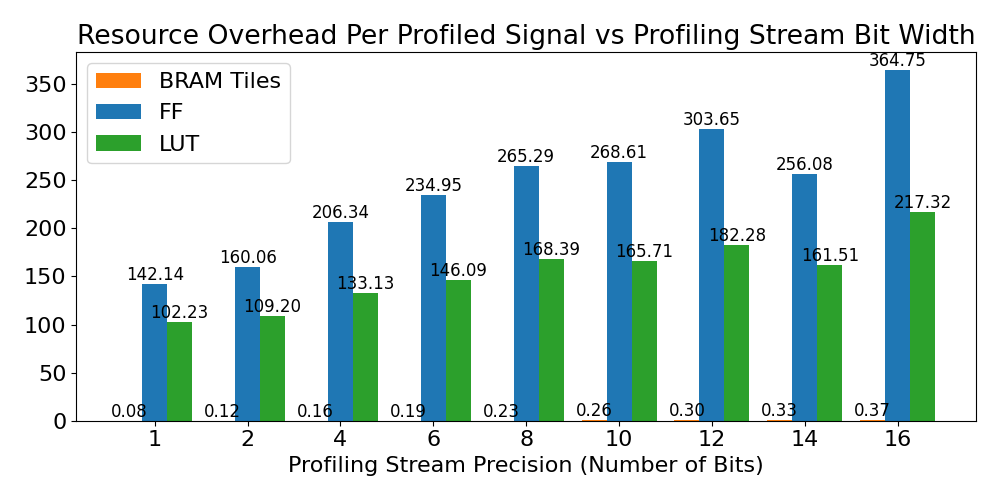}
    \caption{ZCU102, Conv2D stacking. A total of 79 signals are profiled in this RINN for each precision. Data is collected from the Vivado implementation report, with the overhead computed by subtracting the resource requirement of the original (non-profiled) version from the design with profiling structures, and then averaged by the number of concurrently profiled signals. For this FIFO size scenario, bitwidths less than 6 will lead to overflow.}
    \label{fig:resourcevsbitwidth}
\end{figure}

The resource overhead mainly arises from an additional parallel data path that accompanies from the input to the output streams. One inefficiency is that once the profiling data enters the stream, it is repeatedly read and written by subsequent layers. This issue can be alleviated by replacing sufficiently long profiling data streams with dummy ones and introducing intermediate profiling data collection layers that forward the recorded data directly to the end of the profiling path. 

The resource overhead is distributed across individual computation layers. We analyze and compare the overhead using the utilization\_placed.rpt report generated after bitstream compilation in Vivado. For each RINN configuration, we generate two versions of the bitstream: one without the profiling stream and the other with the profiling stream added. The profiled signals range approximately from 0 to more than 200, as shown in Fig.~\ref{fig:resourceoverhead1}.

In Fig.~\ref{fig:resourceoverhead1}, we implemented the RINNs on the ZCU102. The kernel size ranges from 2 to 5, the reshaped layer size ranges from 9 to 36, and the profiled data bitwidth uses ap\_fixed<10,10>, while the data bitwidth uses ap\_fixed<2,1>. We increased the reuse factor as needed to manage resource constraints. In these generated RINNs, the maximum difference between the co-simulation result and the profiled result is 6. The average difference between the co-simulated version and the profiled version is 0.997, while the maximum profiled depth is 66 and the minimum profiled depth is 1.

We also varied the precision of the profiling stream to evaluate how reducing the bitwidth impacts resource usage. We used one RINN stacked with Conv2D layers for this experiment, and the results are shown in Fig.~\ref{fig:resourcevsbitwidth}. Though a bitwidth less than 6 causes overflow in this scenario, it may still be applicable in future cases when profiling other metrics of interest.

\subsection{Case Study: Differences Between CoSim, CSim, and Profiled Results}

\small
\begin{table}[h]
\centering
\begin{tabular}{l>{\centering\arraybackslash}p{1.3cm}>{\centering\arraybackslash}p{1.3cm}>{\centering\arraybackslash}p{1.3cm}>{\centering\arraybackslash}p{1.3cm}}
\toprule
Layer Type & hls4ml default  & CoSim Fifo Size & Profiled Fifo Size & Total Signals\\
\hline
add & 16 & 2 & 1 & 2\\
add & 16 & 2 & 2 & 7\\
add & 16 & 3 & 1 & 1\\
add & 16 & 4 & 3 & 1\\
add & 16 & 10 & 10/11/12 & 1/1/1\\
add & 16 & 11/12/13/14 & 11/12/9/13 & 1/1/1/1\\
add & 16 & 16 & 15 & 10\\
add & 16 & 16 & 16 & 1\\
conv2d & 36 & 10 & 10 & 1\\
conv2d & 36 & 10 & 12 & 7\\
conv2d & 36 & 13 & 9 & 2\\
conv2d & 36 & 13 & 12 & 5\\
conv2d & 36 & 13 & 13 & 1\\
conv2d & 36 & 15 & 15 & 3\\
conv2d & 36 & 26 & 29 & 1\\
clone & 16 & 2 & 1 & 9 \\
relu & 16 & 2 & 1 & 20\\
dense & 1 & 1 & 1 & 1\\
\bottomrule
\end{tabular}
\caption{Profiling result for the RINN used in Fig. \ref{fig:resourcevsbitwidth}, where 79 internal signals were profiled concurrently. The FIFO size refers to the maximum FIFO fullness observed during on-FPGA execution.}
\label{tab:complexitycompare}
\end{table}
\normalsize

We list the profiling results of the RINN on ZCU102 in Table~\ref{tab:complexitycompare}. As shown in the table, the profiled FIFO sizes differ from those in the CoSim version. We believe this discrepancy is related to mutual interference between the profiling process and the data output, which can be observed in an additional state introduced in Listing~\ref{lst:HDL_code}. Further optimization can be achieved by controlling the generation of this extra state during the Vitis HLS synthesis process, through more fine-grained HDL code generation control. Additionally, embedding a profiling library into the Vitis HLS synthesis flow via pragmas is also promising, as demonstrated for latency profiling framework of RealProbe\cite{RealProbe}. More interesting metrics can be integrated into this profiling library with dedicated Verilog code generation, potentially in combination with the SPRING framework. The total on-chip power of the profiled RINN in Fig. \ref{fig:resourcevsbitwidth}, assuming 16 bit profiling precision, increased from 3.568W to 3.667W in total compared to the unprofiled project.

\subsection{Exploring FIFO Size Patterns Across RINN Generation Strategies}
We explore FIFO size patterns by varying several factors, including the complexity of the RINN, the target board, layer types, connection strategy, kernel sizes, filter sizes, the hls4ml reuse factor, and the data bitwidth. The FIFO size refers to the maximum FIFO fullness observed during on-board execution. When referring to the layer type (e.g., Conv2D, Add, Dense), we specifically mean the layer that consumes data from the FIFO.

\begin{figure*}[ht]
    \centering
    \includegraphics[width=\linewidth]{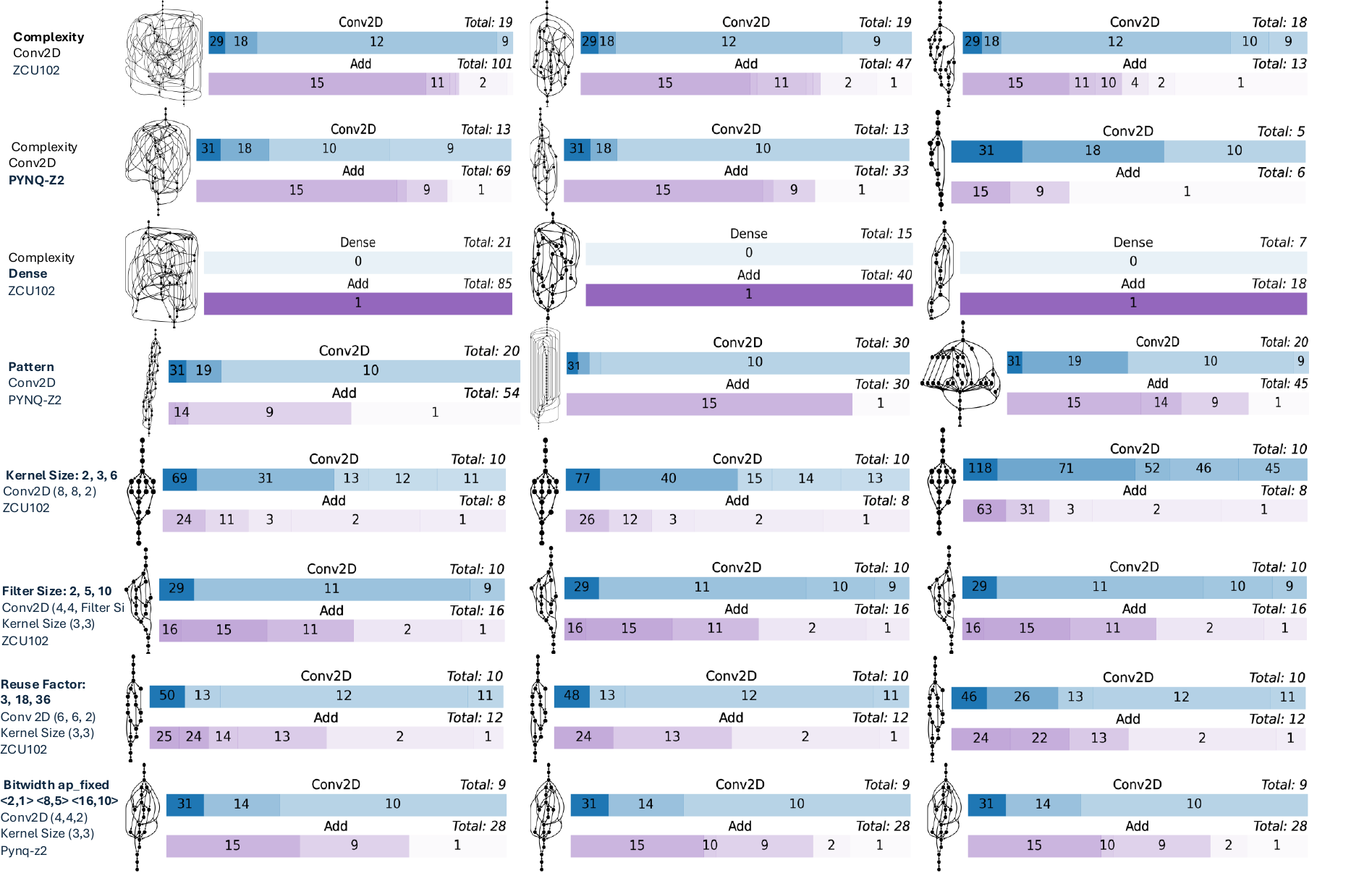}
    \caption{Complexity Influence on FIFO Fullness.}
    \label{fig:complexity}
\end{figure*}

\subsubsection{RINNs Complexity}
We first explored the changes in FIFO size across RINNs with varying complexity. We found that, despite significant variations in network complexity, certain specific FIFO depths consistently emerge. For example, a depth of 29 frequently appears in the first Conv(x,x,1) layer, while depths of 18, 12, and 9 are commonly observed in other Conv(x, x, kernel size) layers. Similar patterns are also encountered in the add layers.

\subsubsection{Differences Across FPGA Devices}

We have experimented with both the ZCU102 and the Pynq-Z2. Similar specific values of FIFO depth continued to appear, though the numbers were slightly different. We deployed the same RINN model on both ZCU102 and Pynq-Z2 separately. Although the Vitis HLS code was exactly the same, the co-simulation and profiling results differed. This may be due to different HDL code being generated for each board. For example, in the matrix multiplication block of the final dense layer, the Pynq-Z2 version uses a register to buffer the output, while the ZCU102 version does not, as seen in the generated Verilog code.

\subsubsection{Variations Across Layer Types}

Instead of stacking Conv2D layers, we randomly connected Dense layers without the need for reshaping or flattening. However, regardless of the network's complexity, the maximum FIFO size for Dense layers remained zero, and the co-simulation FIFO size consistently remained at one. The same behavior was observed on the Pynq-Z2 platform. These findings suggest that FIFO size patterns are highly correlated with the layer types and the way the code is written.

\subsubsection{Variations due to Connection Patterns}

We conducted a preliminary exploration of how FIFO sizes change with different connection strategies occuring within the RINNs. We selected three patterns: short-distance skip connections, long-distance skip connections, and a pattern where most layers connect only to the first few and last few layers, without intermediate connections. Based on this example, we observed that the long-distance pattern resulted in larger FIFO sizes for the Add operations. This opens up opportunities for future exploration.

\subsubsection{Kernel Size of Conv2D}
We changed the kernel size of the Conv2D layer. The reshape layer is set to (8, 8, 2), where 2 is the number of filters. The kernel sizes range from (2, 2), (3, 3), to (6, 6). Based on these examples, we observed that, in general, larger kernel sizes can result in larger FIFO sizes.

\subsubsection{Filter Size of Conv2D}

We varied the filter size, ranging from 2 and 5 up to 10. As observed in the examples, the FIFO size remains mostly unchanged, suggesting that filter size has limited impact on FIFO utilization. In some cases, some FIFO sizes in a Conv2D layer decrease from 11 to 10 as the filter size increases.

\subsubsection{Reuse Factor of \texttt{hls4ml}}

Reuse Factor is a tunable parameter in \texttt{hls4ml}. It determines how many times a multiplier is reused for result calculation within a layer. As shown in Fig.~\ref{fig:complexity}, the reuse factor influences the FIFO size, although the specific trend remains to be explored.

\subsubsection{Bitwidth of the Computation Data Path}

We changed the data bitwidth from ap\_fixed<2,1> and ap\_fixed<8,5> to ap\_fixed<16,10>. The FIFO size remains mostly unchanged, suggesting that bitwidth has limited impact on FIFO utilization. We did observe one case where an add function's FIFO size increased from 9 to 10 when the bitwidth increased. In further runs, using ap\_fixed<4,2> resulted in a FIFO size of 9, while ap\_fixed<6,3> led to a FIFO size of 10.

Based on the above initial observations, we suggest the following strategy: reduce the bitwidth and filter size to the minimum, increase the reuse factor to save resource, and keep the kernel size unchanged. A randomly connected pattern with a small RINN is sufficient to reveal most of the FIFO size candidates needed. After identifying these candidates, larger FIFO sizes can be assigned to long-distance connections to set the initial FIFO sizing.

\section{Conclusion} \label{conclusion}


In this paper, we propose a new profiling framework for HLS-based streaming cores on FPGAs to enable on-board metric verification. The method introduces a parallel profiling stream that flows alongside the main data stream. We apply this approach to RINNs and automate the entire process from RINN generation to on-FPGA metric analysis, using a one-click pipeline built on top of \texttt{hls4ml}. In our implementation, we focus on profiling FIFO size and compare the results against co-simulation data. Our method enables profiling of over 200 internal signals per design, which is challenging for traditional approaches.

Based on the profiled results, we provide insights into how to set initial FIFO sizes during the early stages of ML hardware accelerator development. One limitation is that certain internal logic must be embedded to access specific values of interest, such as the maximum FIFO depth. Additionally, the number of profiled values per signal must be statically known.

Future work includes optimizing the profiling stream by forwarding sufficiently long profiling data directly to the output and minimizing redundant buffering. Simplifying the profiling logic via pragmas and designing dedicated HDL during synthesis can help reduce both data interference and resource overhead. The profiling library can also be extended to include additional metrics such as latency and internal runtime states. These improvements may help ensure safe FPGA operation and offer deeper insights into the hardware behavior of black-box ML models.


\bibliographystyle{IEEEtran}

\bibliography{fpga}
\vspace{12pt}

\end{document}